\documentclass[onecollarge]{svjour2}       % onecolumn "king-size"
\smartqed  % flush right qed marks, e.g. at end of proof
\usepackage{graphicx}
\usepackage{amsmath}
\begin{document}

\title{On the Generalized Langevin Equation for a Rouse Bead in a Nonequilibrium Bath}

\author{Hans Vandebroek         \and
        Carlo Vanderzande %etc.
}


\institute{Hans Vandebroek \at
            Hasselt University, Agoralaan 1, 3590 Diepenbeek, Belgium \\
             % Tel.: +123-45-678910\\
               %  \\
               \email{hans.vandebroek@uhasselt.be} 
           \and
           Carlo Vanderzande \at
           Hasselt University, Agoralaan 1, 3590 Diepenbeek, Belgium and \\
           Katholieke Universiteit Leuven, Department Theoretical Physics, Celestijnenlaan 200D, 3001 Heverlee, Belgium
             % Tel.: +123-45-678910\\
              \email{carlo.vanderzande@uhasselt.be} 
            %  second address
}

\date{Received: date / Accepted: date}

\maketitle

\begin{abstract}
We present the reduced dynamics of a bead in a Rouse chain which is submerged in a bath containing a driving agent that renders it out-of-equilibrium. We first review the generalized Langevin equation of the middle bead in an equilibrated bath. Thereafter, we introduce two driving forces. Firstly, we add a constant force that is applied to the first bead of the chain. We investigate how the generalized Langevin equation changes due to this perturbation for which the system evolves towards a new equilibrium state after some time. Secondly, we consider the case of stochastic active forces which will drive the system to a nonequilibrium state. Including these active forces results in a frenetic contribution to the second fluctuation-dissipation relation, in accord with a recent extension of the fluctuation-dissipation relation to nonequilibrium. The form of the frenetic term is analysed for the specific case of Gaussian, exponentially correlated active forces. We also discuss the resulting rich dynamics of the middle bead in which various regimes of normal diffusion, subdiffusion and superdiffusion can be present. 
\end{abstract}

\section{Introduction}

Complex systems consisting of a very large number of constituents (e.g. colloidal particles suspended in a fluid) can, in principle, be deterministically described by Newton's laws of motion. However, due to the enormous amount of degrees-of-freedom, the resulting set of equations will not be solvable, neither analytically nor numerically. However, one is often only interested in the dynamics of one, or a few, slow variables. We will refer to these variables as ``the system''. One can think of the position of a tagged particle inside a bath of other particles, or the evolution of a reaction coordinate used to describe the folding of a biopolymer. The equation of motion of the slow variable can then be obtained by integrating out the fast variables interacting with the slow one. This approach will produce a closed equation of motion for the slow variable \cite{Zwanzig}. Such an effective equation of motion is commonly known as a generalized Langevin equation. Generally speaking, the generalized Langevin description introduces three distinct terms in the effective dynamics of the system. Namely, an effective potential, a non-Markovian friction force and a non-white effective noise. For a system in an equilibrium environment, the latter two are connected by the (second) fluctuation-dissipation relation.

The simplest example of a generalized Langevin equation is the original Langevin equation introduced by Paul Langevin in $1908$ \cite{PLangevin}. It gives the reduced dynamics of a Brownian particle in a fluid. Equating the Newtonian inertial term are only two forces, a Stokesian friction force and a thermal random noise. Both these forces are Markovian processes, implying that they are instantaneous effects, not influenced by the system's past. This, rather simple, model is ideal to describe the diffusion of a small particle in a viscous fluid. That is because the time scale of collisions between the system and fluid particles, which correlates them, is very short.

\begin{figure*}
\center
  \includegraphics[width=0.5\textwidth]{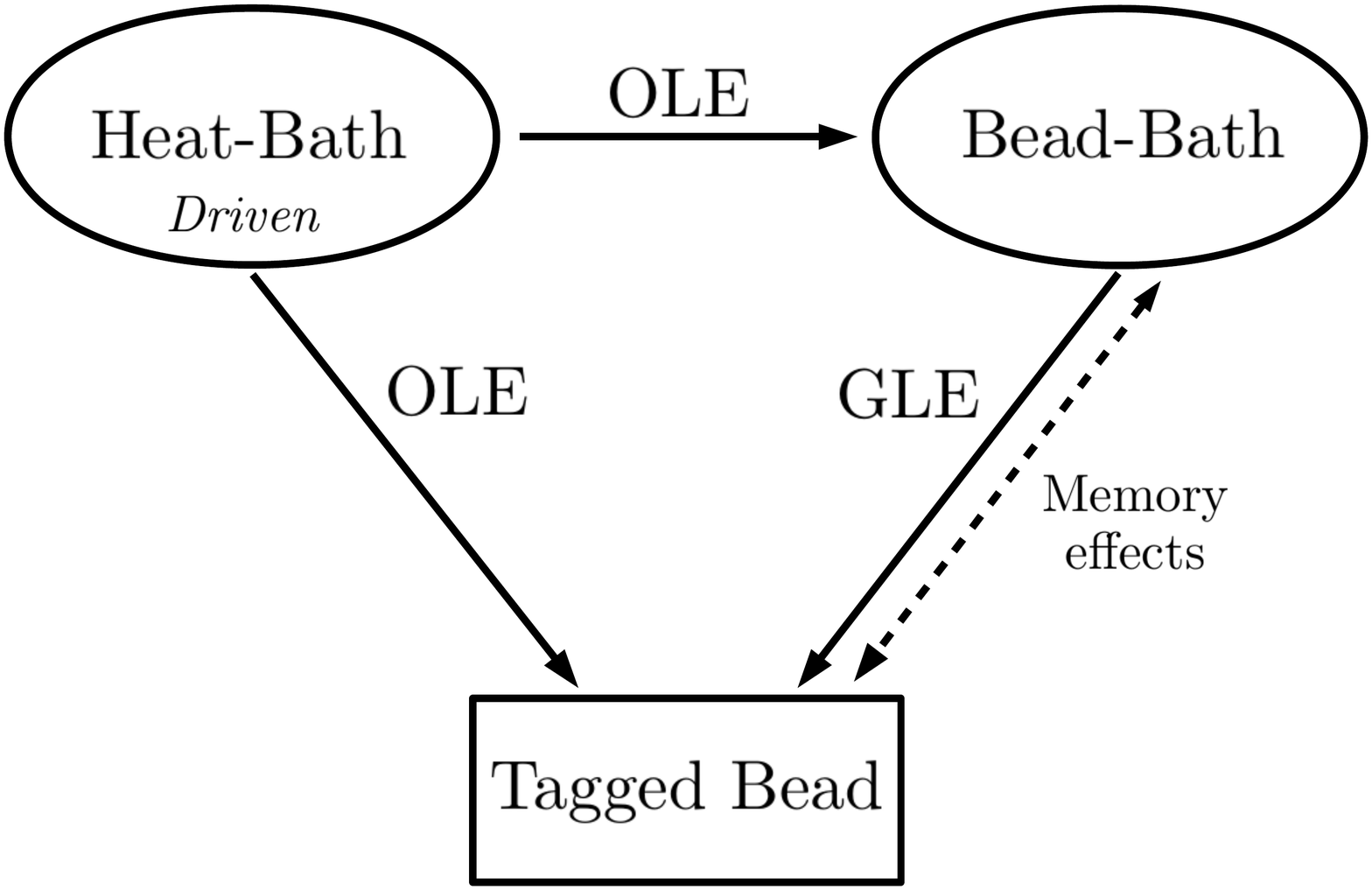}
\caption{Schematic representation of the applied description to acquire the reduced dynamics resulting from the different baths. OLE stands for ordinary Langevin equation and GLE stands for generalized Langevin equation.}
\label{fig:model}       
\end{figure*}

The effective dynamics we will investigate, are those of a ``tagged bead'' in a Rouse chain. A Rouse chain is comprised of masses (or beads) linearly connected by harmonic springs and submerged in a viscous fluid. It is the simplest model for the dynamics of a polymer \cite{RouseE,Doi}. One of these beads will function as our system for which we construct its generalized Langevin equation. The dynamics of our system is influenced by two baths. First, we have the interaction with the particles of the fluid, which we refer to as the heat-bath. We apply the original Langevin equation to describe the reduced dynamics resulting from this bath. This implies that no memory effects will arise. Naturally, the heat-bath not only couples to the system but also to all other beads. Second, because the tagged bead is part of a chain, its dynamics depend on the evolution of the rest of the chain, which we call the bead-bath. Projecting the degrees-of-freedom of the bead-bath on the system will yield a generalized Langevin equation that represents the reduced dynamics of the system originating from the interaction with the chain. The bead-bath will therefore impose memory effects, which was to be expected due to the interconnected nature of the tagged bead and all other beads. Figure \ref{fig:model} gives a schematic representation on how the reduced dynamics of the tagged bead are acquired from the two baths.

For an equilibrated heat-bath the reduced dynamics of the tagged bead was already derived by D. Panja \cite{Panja}. We will first give a short review of his findings. Our work extends these results to a heat-bath which is not in equilibrium but possesses some driving agent. We investigate two specific driving agents: a constant force on the first bead and nonconservative active forces which will, in contrast with the constant force, pull the system away from equilibrium indefinitely. 

A Rouse chain is widely known as a model that describes the dynamical properties of a polymer. A bead in the chain represents a rigid group of monomers. Many experiments study the diffusion of a tagged monomer when the polymer is surrounded by the complex cellular environment \cite{Weber,Bronstein}. Knowing the effective dynamics of such a monomer is therefore crucial if one want to interpret the observed diffusion, which often is anomalous. Furthermore, the cellular environment is certainly not in equilibrium. The nonequilibrium processes we study are therefore inspired by biological phenomena. The constant force on the first bead is a model for the process where molecular motors drag a polymer through the cellular environment in a well defined direction. This dragging takes, for example, place during cell division where the chromosomes are pulled apart by the mitotic spindle \cite{Phillips}. The active forces on the other hand model the overall non-thermal fluctuations inside cell, resulting from the action of molecular motors on the cytoskeleton. The precise characterisation of these active forces inside a cell or amidst {\it in vitro} cytoskeletons is a matter of current research \cite{Robert,Levine,Fodor}. We will assume the forces to be Gaussian as was recently shown to be true in artificial actin-myosin networks in a regime where the number of myosin motors is small \cite{Sonn-Segev}.

\section{From Rouse Model to Generalized Langevin Description}

A Rouse chain is a simple model for a polymer devised by Prince E. Rouse in $1953$ \cite{RouseE}. It comprises of $N$ beads with mass $m$ linearly connected by harmonic springs with spring constant $k$. This chain is submerged in a viscous heat-bath at temperature $T$ and with friction coefficient $\gamma$. The motion of the $n^{th}$ bead, with $n=1,2,\dots,N$, is given by the original Langevin equation. The position of this bead is given by vector $\vec{R}_n$ and thus we have
\begin{equation}
m\, \ddot{\vec{R}}_n(t) = - k\,\Big(2\vec{R}_n(t)-\vec{R}_{n-1}(t)-\vec{R}_{n+1}(t)\Big) -\gamma\,\dot{\vec{R}}_n(t)  + \vec{\xi}_{T,n}(t). \label{Rouse}
\end{equation}
In order to apply this formula to all beads, one introduces two ghost beads at $n=0$ and $n=N+1$ whose positions satisfy $\vec{R}_0=\vec{R}_1$ and $\vec{R}_{N+1}=\vec{R}_{N}$. The two last terms on the right-hand side of equation (\ref{Rouse}) represent the (reduced) interaction of the $n^{th}$ bead with the heat-bath. They are, respectively, the Stokesian friction force and thermal agitation. These processes are connected through the fluctuation-dissipation relation which states that $\langle \vec{\xi}_{T,n}(t) \cdot\vec{\xi}_{T,m}(t')\rangle=6\gamma k_BT\delta(t-t')\delta_{n,m}$, with $k_B$ the Boltzmann constant. Furthermore, these stochastic forces are assumed to be Gaussian with zero mean. The first term on the right-hand side originates from the harmonic interaction of the springs. If we assume to be in a low Reynolds number regime, we can apply the overdamped limit to this equation, i.e. $m/\gamma\ll1$. This implies that the inertial term on the left-hand side of equation (\ref{Rouse}) drops out. 

To acquire the reduced dynamics of a particular tagged bead, one can eliminate all other degrees-of-freedom from the set of equations of motion that (\ref{Rouse}) entails. This can be done by a change of variables to normal coordinates (see \cite{Doi}). This effectively projects the (slow) dynamics of all beads on the dynamics of the tagged bead. It leads to an equation of motion known as the generalized Langevin equation of the system (i.e. the tagged bead). This calculation was first performed in \cite{Panja} in the limit where $n$ is taken as a continuous variable. We give results for a discrete chain. The detailed derivation can be found in chapter $5$ of \cite{thesis}. Here, we will provide a short overview of its expression and implications for the middle bead as the system. 

Consider a Rouse chain of $N=2M+1$ beads, the reduced dynamics of the position $\vec{r}(t)=\vec{R}_{M}(t)$ of the middle bead obeys the following generalized (overdamped) Langevin equation \cite{Panja}
\begin{equation}
\gamma \, \dot{\vec{r}}(t) = \vec{\xi}_T(t)  - \int_0^t d\tau\, K(t-\tau)\,\dot{\vec{r}}(\tau) + \vec{\Phi}(t). \label{reddyn}
\end{equation}  
The term on the left-hand side and the first term on the right hand side of this equation are the friction and thermal noise acting on the middle bead, which already appear in (\ref{Rouse}) and which have not been integrated out.  The other two forces on the right-hand side of the equation of motion originate from the complex interaction with the remaining $2M$ beads. The convolution integral represents a non-Markovian friction on the middle bead. The past velocity profile of the bead influences the present friction, this influence is characterized by a memory kernel $K(t)$. The memory kernel is a large sum over decaying exponentials 
\begin{equation}
K(t) = \frac{8k}{N}\,\sum_{p=1}^{M}\cos^2\left( \frac{(2p-1)\pi}{2N} \right) \exp(-t/\tau_p). \label{eqKernel}
\end{equation}
The characteristic times of the exponentials are given by $\tau_p=\gamma N^2/k\pi^2(2p-1)^2$ for large $N$. The slowest time $\tau_1 \sim N^2$ is referred to as the Rouse time and it is physically related to the time for the polymer to diffuse over its own radius of gyration. It can be shown that for large $N$ this kernel has the following approximate form $K(t)\approx\sqrt{4\gamma k/\pi}\,\exp(-t/\tau_1)\,t^{-1/2}$, i.e. it displays an initial power-law decay followed by a decaying exponential. 

The last term in the equation of motion is the effective noise. It is Gaussian distributed with zero mean. Its correlation is connected to the non-Markovian friction through the fluctuation-dissipation relation. We have
\begin{equation}
\langle \vec{\Phi}(t) \cdot \vec{\Phi}(t')  \rangle = 3k_BTK(|t-t'|).
\end{equation}
From this equation it is clear that the effective noise is not white but correlated through time, i.e. it is coloured noise. Just as previous velocities of the system can influence its present friction, so will effective noise from the past have effect on the present effective noise. Both effects are dictated by the memory kernel. Because both noise terms, thermal and effective, obey the fluctuation-dissipation relation, the system will surely reach thermal equilibrium. Notice that in this case, the bead-bath does not produce an effective potential on the middle bead.

In the following two sections, we will assume that the heat-bath in addition possesses some driving force that will pull our system away from equilibrium. First, we look at a constant force which is applied to the first bead. Because this force is conservative, the system will regain thermal equilibrium after some characteristic time. But during the transient regime where the whole chain responds to the new force the system will not be in equilibrium. Second, we assume the heat-bath exerts, apart from the thermal random forces, active random forces.

\section{Constant Force}

Applying a constant force to the first bead will add an extra term to equation (\ref{Rouse}). This term is given by $\vec{f}_n(t)=fH(t)\delta_{n,0}\vec{e}_x$. The strength of the force is determined by $f$ and the Heaviside function $H(t)$ ensures that the force is turned on at $t=0$. The arbitrary direction of the force is in the positive $x$-direction. The effect of such a term on the motion of a bead was already investigated by T. Sakaue \cite{Sakaue}. We will however approach the problem from a slightly different angle. 

\begin{figure*}
\center
  \includegraphics[width=0.5\textwidth]{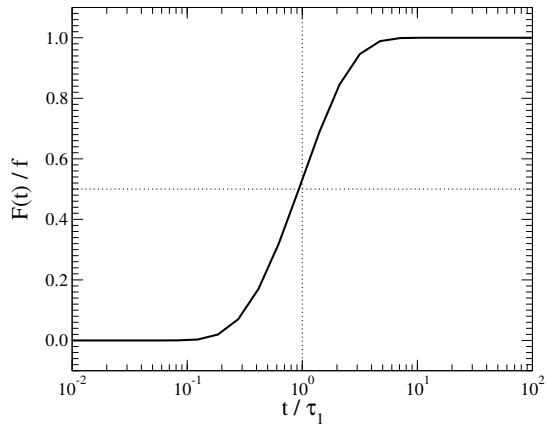}
\caption{Lin-log plot of the ratio of the effective force to the constant force as a function of time. We have $F(t)=|\vec{\mathcal{F}}(t)|$. The vertical dotted line indicates the longest characteristic time $\tau_1$. The horizontal dotted line is drawn at $F/f=1/2$. The number of beads in the chain is $N=1025$, also $\gamma=1$ and $k=3$.}
\label{fig:constant}       
\end{figure*}

Taking into account this force during the elimination of the bead-bath variables leads to the following reduced dynamics of the middle bead  
\begin{equation}
\gamma \, \dot{\vec{r}}(t) = \vec{\xi}_T(t)  - \int_0^t d\tau\, K(t-\tau)\,\dot{\vec{r}}(\tau) + \vec{\Phi}(t) + \vec{\mathcal{F}}(t).
\end{equation}
The new term on the right-hand side of this equation of motion represents the gradual increase of the constant force's influence on the middle bead. This effective force is given by (see \cite{thesis} for the derivation)
\begin{equation}
\vec{\mathcal{F}}(t) = \frac{2f}{N}\sum_{p=1}^{M} (-1)^p\cos\left( \frac{(2p-1)\pi}{2N} \right) \cot\left( \frac{(2p-1)\pi}{2N} \right) \Big[ \exp(-t/\tau_p) -1\Big]H(t)\, \vec{e}_x. \label{efffor}
\end{equation}
It is clear that when the elapsed time is much shorter than the smallest characteristic time, i.e. $t\ll\tau_M$, the effective force on the middle bead is essentially zero. The constant force has not yet had time to propagate through the chain to reach the middle bead, leaving this bead unaware of its presence. So for such early times we retrieve the equilibrium generalized Langevin equation (\ref{reddyn}), implying that the system remains thermalized. When the force does reach the tagged bead, the system will undergo a transient phase where the effective force on it will build up. We refer to \cite{Sakaue} and \cite{us1}, for a discussion on the resulting time-evolution of the system. For long times, $t\gg\tau_1$, the effective force takes on a constant value, resulting again in a thermalized system. This constant value should be equal to the original constant force since it has fully propagated the chain. We indeed find this to be true for $N\gg1$
\begin{equation}
\vec{\mathcal{F}}(t\gg\tau_1) = -\frac{2f}{N}\sum_{p=1}^{M}(-1)^p\cos\left( \frac{(2p-1)\pi}{2N} \right) \cot\left( \frac{(2p-1)\pi}{2N} \right) \vec{e}_x \\ = -\frac{2f}{N}(-M) \vec{e}_x = f\, \vec{e}_x.
\end{equation}
Figure \ref{fig:constant} shows the time-evolution of the effective force's magnitude compared to that of the constant force. It is a numerical evaluation of relation (\ref{efffor}). As expected, the effective force is, for early times, zero and for long times it is equal to $f$. It reaches about half of its final value at the longest characteristic time $\tau_1$. In fact, in \cite{us1} it was shown that the effect of the force diffuses through the chain, so that the time to reach the middle bead is of the order $(N/2)^2$, i.e. this time is of the order of the Rouse time. 

\section{Active Forces}

We now assume that in addition to thermal noise, the beads feel active random forces. These could arise from the interaction of the polymer with active Brownian particles as was investigated in recent simulations \cite{Kaiser,Shin}. Or they could represent active forces on a biopolymer immersed in a cell or in an artificial cytoskeleton \cite{Brangwynne}. Then we must add a stochastic term, $\vec{\xi}_{A,n}(t)$, to the equations of motion (\ref{Rouse}). We first consider the general case where this new random force is Gaussian distributed with zero mean and a correlation that is a function of the absolute difference in time and uncorrelated between beads
\begin{equation}
\langle \vec{\xi}_{A,n}(t) \cdot \vec{\xi}_{A,m}(t') \rangle = A(|t-t'|)\delta_{n,m}(1-\delta_{n,N/2}). \label{actcorr}
\end{equation} 
The second delta assumes that the active random forces do not couple to the tagged bead. Extending the model to one where they do couple is trivial and does not yield conceptually different results.  After eliminating the degrees-of-freedom of the chain, one finds that the expression of the generalized Langevin equation (\ref{reddyn}) still holds. There is no change in the friction term, the kernel $K(t)$ is still given by (\ref{eqKernel}). However, the correlation of the effective noise will be seriously affected. In a steady state regime, where $t$ and $t'$ are large, the correlation has the following expression
\begin{equation}
\langle \vec{\Phi}(t) \cdot \vec{\Phi}(t')  \rangle_{neq} = 3k_BT \, \Big( K(|t-t'|) + K_+(t,t') \Big). \label{frencon}
\end{equation} 
The derivation of this relation is lengthy, but involves no special techniques. The details can be found in \cite{thesis}. 

As can be seen, the correlation of the effective noise now comprises of two terms, the entropic memory kernel $K$ from the equilibrium bath and a new memory kernel $K_+$. The effective noise is, however, still Gaussian with zero mean. From this expression we clearly see that the effective noise no longer obeys the fluctuation-dissipation relation, because no corresponding dissipative term exist for $K_+$. Disobeying this relation results in a drift away from thermal equilibrium, hence the subscript ``$neq$'' which indicated the nonequilibrium character of this correlation. This result is an explicit example of the general form of the second fluctuation-dissipation theorem out of equilibrium as recently studied by C. Maes \cite{Maes,MaesS}. The nonequilibrium effective noise now consists of a term that obeys the fluctuation-dissipation relation and a term that doesn't. Following Maes, we call $K_+$ a  frenetic contribution. It is given by 
\begin{equation}
K_+(t,t') = \frac{8k^2}{3\gamma^2k_BTN} \sum_{p=1}^{M} \mathcal{S}_p^2 \int_0^t d\tau \int_0^{t'} d\tau' \, A(|t-t'|)\,  \exp(-(t-\tau)/\tau_p) \, \exp(-(t'-\tau')/\tau_p),
\end{equation}
where $\mathcal{S}_p=\sin\left((2p-1)\pi/N\right)$.

The above expression for the frenetic contribution can be solved for some specific forms of the correlation $A$ of the new random force. Here we will present the case where the driving force is characterized by an exponential correlation because of its relevance for a polymer in a bath of active Brownian particles or in an actin-myosin network \cite{Levine}. We assume
\begin{equation}
A(|t-t'|) = 3C\exp(-|t-t'|/\tau_A). \label{corrA}
\end{equation}
These active random forces can be understood as directionally persistent forces of characteristic strength $\sqrt{C}$ and persistence time $\tau_A$. This time gives the average time over which these random forces maintain their direction, after which they choose an uncorrelated new direction. After a lengthy, yet elementary, calculation one finds
\begin{equation}
K_+(|t-t'|) = \frac{8Ck^2}{\gamma^2k_BTN}\sum_{p=1}^M \frac{\tau^2_p\,\mathcal{S}_p^2}{1-(\tau_p/\tau_A)^2}\, \left[  \exp(-|t-t'|/\tau_A)  - \left(\frac{\tau_p}{\tau_A}\right) \exp(-|t-t'|/\tau_p) \right]. \label{K++}
\end{equation}
Just as the equilibrium memory kernel, $K_+$ is a sum over exponentials and only a function of the absolute difference in time. We can approximate this expression when we allow some assumptions on the active forces characteristics to be made. 

\subsection{Large persistence time}

When $\tau_A\gg\tau_1$, we find the following simple form for the frenetic contribution
\begin{equation}
K_+(|t-t'|) = \frac{C\,N}{k_BT}\,\exp(-|t-t'|/\tau_A). \label{largeK}
\end{equation} 
We used here that $\sum_{p=1}^M (\tau_p\mathcal{S}_p)^2 = \gamma^2N^2/8k^2$ for large $N$. This expression can be understood in the following way: in the previous section, on the constant  force, we found that a force on a bead needs a duration of order $\tau_1$ to fully reach the middle bead (see figure \ref{fig:constant}). Because the persistence in a particular direction of these active forces is indeed much longer than $\tau_1$, they act like a constant force for a considerable amount of time. Therefore, all beads fully transpose the active force on them to the middle bead, yielding $N$ times equation (\ref{corrA}). The nonequilibrium effective noise subsequently becomes
\begin{equation}
\langle \vec{\Phi}(t) \cdot \vec{\Phi}(t')  \rangle_{neq} = 3k_BT \,K(|t-t'|) + NA(|t-t'|).
\end{equation}
This result is general for any function $A$ that decays with a typical time scale $\tau_A \gg \tau_1$. 
One can interpret this result as two effective noises. The first are the equilibrium thermal random forces, governed by memory kernel $K(t)$. The second are the original active forces that act on every bead, but enhanced by a factor $N$.

\subsection{Small persistence time}

When we take $\tau_A\ll\tau_M$ in equation (\ref{K++}), we can do the following
\begin{equation}
K_+(|t-t'|) = - \frac{8Ck^2}{\gamma^2k_BTN}\sum_{p=1}^M \tau^2_A\,\mathcal{S}_p^2 \left[  \exp(-|t-t'|/\tau_A)    - \left(\frac{\tau_p}{\tau_A}\right) \exp(-|t-t'|/\tau_p) \right].  
\end{equation}
Using the definition of the equilibrium memory kernel (\ref{eqKernel}) and applying $\sum_{p=1}^M\mathcal{S}_p^2\approx N/4$ for large $N$, results in
\begin{equation}
K_+(|t-t'|) = -\frac{2\tau_A^2Ck^2}{\gamma^2k_BT}\,\exp(-|t-t'|/\tau_A) + \frac{\tau_A\,C}{\gamma k_BT}\,K(|t-t'|).
\end{equation}
The first term in this expression can be neglected since the persistence time $\tau_A$ is very small. The nonequilibrium effective noise becomes
\begin{equation}
\langle \vec{\Phi}(t) \cdot \vec{\Phi}(t')  \rangle_{neq} = 3k_BT_* \,K(|t-t'|), \label{effnoilow}
\end{equation}
with $k_BT_*=k_BT+\tau_AC/\gamma$. We thus find an effective noise that has an equilibrium expression but with a higher effective temperature than the case where no active forces are present. The system will therefore appear to thermalize under this new temperature. The reason for this behaviour is because the low-persistence-time active forces mimic the thermal noise $\vec{\xi}_{T,n}$ on the beads. Using a well known representation of the Dirac delta function
\begin{equation}
\delta(x) = \lim_{\epsilon\rightarrow 0} \frac{1}{2\epsilon}\,\exp(-|x|/\epsilon),
\end{equation}
the correlation of the active forces (i.e. equation (\ref{actcorr}) together with equation (\ref{corrA})) indeed becomes approximately equal to $6\tau_AC\delta(t-t')\delta_{n,m}$. The two Gaussian white noises, thermal and active, can thus be combined into one $\vec{\xi}_n=\vec{\xi}_{T,n}+\vec{\xi}_{A,n}$. Naturally, this new random force is also Gaussian distributed with zero mean. Its correlation is
\begin{equation}
\langle \vec{\xi}_n(t) \cdot \vec{\xi}_m(t') \rangle = 6\gamma \left[ k_BT + \frac{\tau_A\,C}{\gamma}\right]\delta(t-t')\delta_{n,m}.
\end{equation}
When using this noise in equation (\ref{Rouse}) instead of $\vec{\xi}_{T,n}$, it is clear that we can redo the equilibrium calculations, but assuming the effective temperature $T_*$. This will also yield the nonequilibrium effective noise (\ref{effnoilow}) derived above.

\begin{figure*}
\center
  \includegraphics[width=0.5\textwidth]{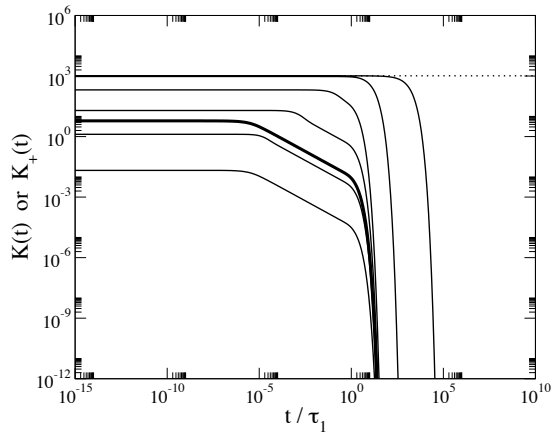}
\caption{Log-log plot of the two kernels as a function of time for $N=1025$. The solid black line represents $K(t)$. The other thinner lines are $K_+(t)$ with $\tau_A/\tau_1=10^{-7}$, $10^{-5}$, $10^{-3}$, $10^{-1}$, $10^{1}$ and $10^{3}$. The higher the value of $\tau_A/\tau_1$, the higher the line lies. The other parameters are $k_BT=\gamma=C=1$ and $k=3$. The straight dotted line is drawn at $CN/k_BT$.}
\label{fig:kernel}       
\end{figure*}

\begin{figure*}
\center
\includegraphics[width=0.45\textwidth]{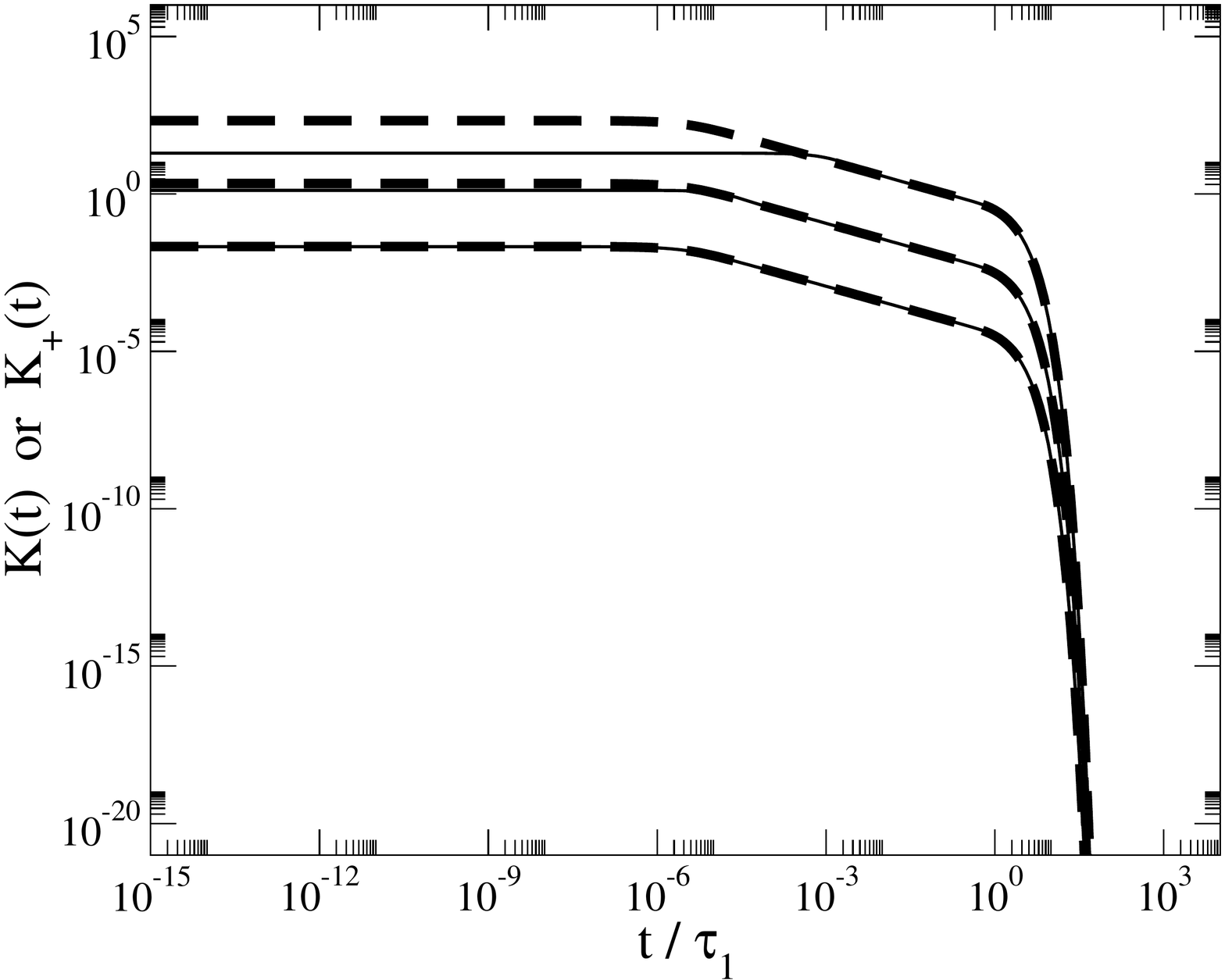} 
\includegraphics[width=0.45\textwidth]{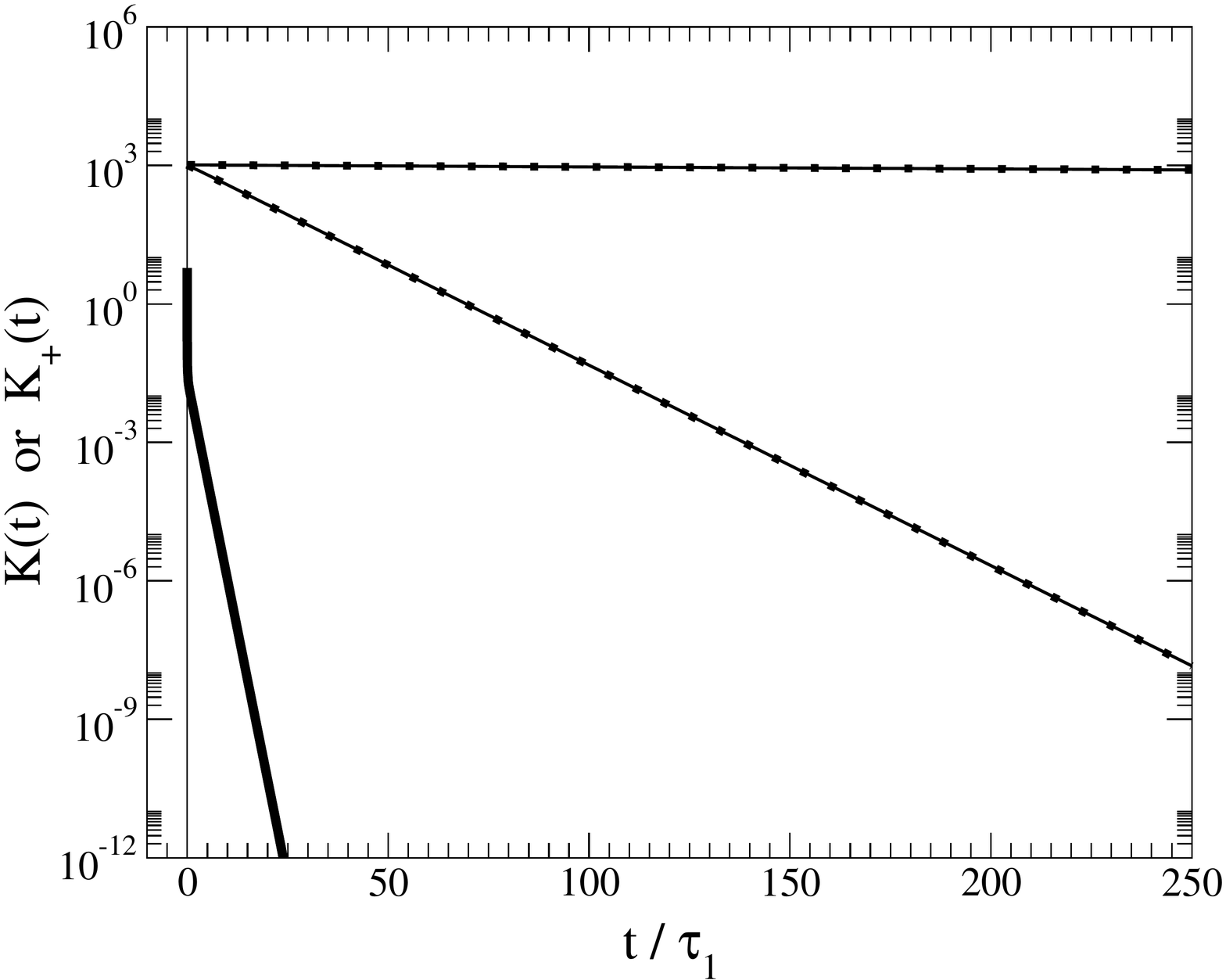}
\caption{{\em Left:} Log-log plot that shows the same three $K_+(t)$ curves from figure \ref{fig:kernel}, with $\tau_A/\tau_1=10^{-7}$, $10^{-5}$ and $10^{-3}$. Also shown in dashed lines are the weighted equilibrium memory kernels, i.e. $(\tau_AC/\gamma k_BT)K(t)$. {\em Right:} Log-lin plot of the two upper $K_+(t)$ curves from figure \ref{fig:kernel}, with $\tau_A/\tau_1=10^1$ and $10^3$. Also shown are $K(t)$ (solid black line) and equation (\ref{largeK}) (dotted lines).}  
\label{fig:kernel1}     
\end{figure*}

Figure \ref{fig:kernel} shows the numerical evaluation of $K(t)$ and $K_+(t)$, equations (\ref{eqKernel}) and (\ref{K++}) respectively. For very early times, the equilibrium memory kernel is constant. Thereafter, for times up to $\tau_1$, this kernel shows a power-law decay after which it crosses over to an exponential decay, as was mentioned before. The behaviour of the frenetic contribution correctly shows a dependency on the value of $\tau_A$. In the left of figures \ref{fig:kernel1} we demonstrate that for small persistence time $\tau_A$, the frenetic contribution is indeed equal to the equilibrium memory kernel multiplied by $\tau_AC/\gamma k_BT$. For large persistence time $\tau_A$ on the other hand, the right figure in \ref{fig:kernel1} shows that $K(t)$ is not exponential for early times while the frenetic contribution is for all times, corresponding nicely with equation (\ref{largeK}). The maximum value of the frenetic contribution never exceeds $CN/k_BT$, which figure \ref{fig:kernel} clearly shows.

\section{Superdiffusive motion of the tagged bead}
To find the time-evolution of the middle bead, one should solve equation (\ref{reddyn}) using the nonequilibrium version of the fluctuation-dissipation relation (\ref{frencon}). Since the equation is linear, this can easily be done using Laplace transform methods. Alternatively, one could solve the set of equations of motion (\ref{Rouse}) with the inclusion of active forces. This alternative derivation was already performed in \cite{us}. In that paper, the more general case of a Rouse chain moving in a viscoelastic environment and in the presence of active forces was studied. Little attention was paid to the viscous limit. 

Using the results of \cite{us} or by direct solution of (\ref{reddyn}) it is possible to obtain the mean squared displacement $\Delta_m^2(t) \equiv\langle (\vec{r}(t)-\vec{r}(0))^2 \rangle$ of the middle bead for the case of exponentially correlated active noise. The result is
\begin{equation}
\begin{split}
\Delta^2_m(t) = \,\,\, &\frac{6}{\gamma N}\,\left(k_BT+\frac{\tau_A\,C}{\gamma}\right)\,t  + \frac{12\,k_BT}{\gamma N} \sum_{p=2,even}^{N-1}\tilde{\tau}_p\, \bigg(1-\exp\big(-t/\tilde{\tau}_p\big)\bigg) + \frac{6\,\tau_A^2\,C}{\gamma^2N}\, \bigg(\exp\big(-t/\tau_A\big)-1 \bigg) \\ +&\frac{12\,C}{\gamma^2N}\sum_{p=2,even}^{N-1} \frac{\tau_A\,\tilde{\tau}^2_p}{\tau_A-\tilde{\tau}_p}\, \Bigg[ \frac{\tau_A}{\tau_A+\tilde{\tau}_p}\,\bigg(1-\exp\big(-(\tau_A^{-1}+\tilde{\tau}^{-1}_p)t\big) \bigg) +\frac{1}{2}\,\bigg(\exp\big(-2t/\tilde{\tau}_p\big) -1\bigg) \Bigg]
\end{split}
\label{mondif}
\end{equation}
with $\tilde{\tau}_p=\tau_{(p+1)/2}$. 

\begin{figure*}
\center
\includegraphics[width=0.5\textwidth]{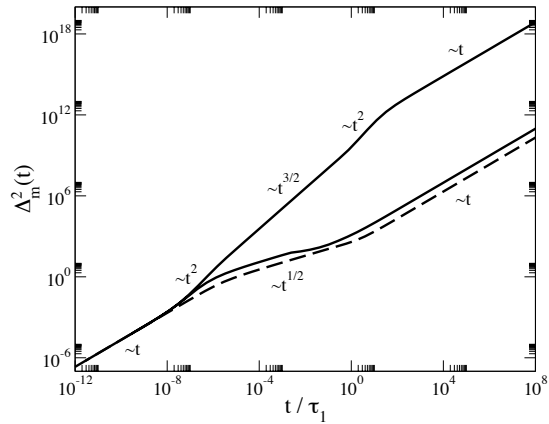} 
\caption{Mean squared displacement of the middle bead in a Rouse chain in absence of active forces (dashed line) and in presence of active forces with small persistence time (lower full line $\tau_A/\tau_1=10^{-7}$) and large persistence time (upper full line $\tau_A/\tau_1=10$). Also indicated are the different regimes of diffusive, subdiffusive and superdiffusive behaviour.}  
\label{msd}     
\end{figure*}

In absence of active forces ($C=0$) only two terms survive. It is well known that for that case there are three time regimes \cite{Doi,Panja}. For $t < \tau_M$, the bead doesn't feel the effect of the neighbouring beads yet, and it diffuses, i.e. $\Delta_m^2(t) \sim t$. For $t > \tau_1$ the bead follows the diffusion of the chain as a whole, i.e. we have again $\Delta_m^2(t) \sim t$ but with a diffusion constant that is a factor $N$ smaller. Finally, in the intermediate time regime $\tau_M < t < \tau_1$ the bead feels that it is inside a chain leading to a subdiffusion with an exponent $1/2$, i.e. $\Delta_m^2(t) \sim t^{1/2}$. This behaviour is shown in Fig. \ref{msd} (dashed line).

In the presence of active forces the motion is more complicated. We first describe the motion of a free particle in presence of thermal and exponentially correlated active forces \cite{thesis,us2}. After a short initial period in which the particle diffuses, it will perform a superdiffusion with an exponent $2$, i.e. $\Delta_m^2(t) \sim t^2$, for the period in which the active forces are persistent, i.e. $t < \tau_A$. For $t > \tau_A$, the free particle will diffuse again. If the particle is part of a chain we expect that, as was the case in absence of active forces, its motion will be modified in the time regime $\tau_M < t < \tau_1$. In fact, in \cite{us} we have shown that the bead will again superdiffuse but with an exponent $3/2$. 

We can now envisage two scenario's. In the first one, $\tau_A > \tau_1$. Then the bead will first move as a free particle, then feel the presence of the other beads, and then will again move as a free particle but with an amplitude that is $N$ times smaller. This type of behaviour indeed shows up if we plot $\Delta_m^2(t)$ as a function of $t$ using (\ref{mondif}) for $\tau_A=10 \tau_1$ (Fig. \ref{msd}, upper full line). If on the other hand the active forces act only for times $\tau_A < \tau_1$ the bead will show the behaviour described above only up to $\tau_A$ after which it moves like a bead in a chain with only thermal noise. For example, in Fig. \ref{msd} (lower full line) we show the motion of the bead when $\tau_A= 10^{-7} \tau_1$. The active forces will lead to a short superdiffusive regime ($\Delta_m^2(t) \sim t^2$) after which the bead has subdiffusive, followed by ordinary diffusive motion. This motion looks very much like that in absence of active forces, the only difference is a somewhat larger prefactor. This is in agreement with what we learned in the previous section, i.e. for $\tau_A \ll \tau_1$, where the behaviour of the bead could be described in terms of an effective temperature.

\section{Conclusions}

The general description for the dynamics of a slow variable interacting with a nonequilibrium bath was provided in \cite{Maes,MaesS}. Here, as a concrete solvable example, we studied the equation of motion of a bead in a Rouse chain with a constant force on the first bead and in an active environment. When the driving agent consisted of a constant force on the first bead, we showed how this force propagates through the chain, increasing its influence on the middle bead. This introduced the middle bead to a transient regime after which it regained thermal equilibrium. When the active forces acted as the driving agent, we found that the fluctuation-dissipation relation picks up a second memory term. Apart from the equilibrium entropic memory kernel, a frenetic contribution is added to the correlation of the effective noise. This complies with the general result in \cite{Maes}. For exponentially correlated active forces, we find that when the persistence time is large, the frenetic contribution is $N$ times the active correlation on a particular bead. When the persistence time is small, the system can be described with equilibrium reduced dynamics at an effective temperature $T_*=T+(\tau_AC/\gamma k_BT)$.  
These two regimes lead to two different possible motions for the tagged bead. For persistence times that are large in comparison with the Rouse time, various regions of diffusion and superdiffusion are expected. If on the other hand the persistence time is much smaller than the Rouse time, the only effect of the active forces will be that the motion in absence of a active forces (diffusive, subdiffusive, diffusive) will be observed but the amplitude of that motion will be somewhat larger, an effect that can be seen as coming from a higher effective temperature.

\end{document}